\documentclass{article}
\usepackage{spconf,amsmath,graphicx}
\usepackage{mwe} 
\usepackage{upgreek}
\usepackage[skip=2pt]{caption}
\usepackage{multirow}

\graphicspath{{figures/}}

\usepackage{hyperref}
\hypersetup{linkbordercolor={1 1 1},citebordercolor={0 0
0},urlbordercolor={0 1 0},pdfborder={0 0 0},breaklinks=false}
\usepackage{makecell}

\def\x{{\mathbf x}}

\newcommand{\rpm}{\raisebox{.0ex}{$\scriptstyle\pm$}}

\newcommand{\comment}[1]{}

\title{BAYESIAN OPTIMIZATION OF 2D ECHOCARDIOGRAPHY SEGMENTATION}

\twoauthors
 {Tung Tran, Joshua V. Stough\sthanks{Corresponding author: {\tt joshua.stough@bucknell.edu}}}
	{Bucknell University\\
	Computer Science\\
	Lewisburg, PA}
 {Xiaoyan Zhang, Christopher M. Haggerty}
	{Geisinger\\
	Translational Data Science and Informatics\\
	Danville, PA}
\begin{document}
%
\maketitle
%


\begin{abstract}
Bayesian Optimization (BO) is a well-studied hyperparameter tuning technique that is more efficient than grid search for high-cost, high-parameter machine learning problems. Echocardiography is a ubiquitous modality for evaluating heart structure and function in cardiology. In this work, we use BO to optimize the architectural and training-related hyperparameters of a previously published deep fully convolutional neural network model for multi-structure segmentation in echocardiography. In a fair comparison, the resulting model outperforms this recent state-of-the-art on the annotated CAMUS dataset in both apical two- and four-chamber echo views. We report mean Dice overlaps of 0.95, 0.96, and 0.93 on left ventricular (LV) endocardium, LV epicardium, and left atrium respectively. We also observe significant improvement in derived clinical indices, including smaller median absolute errors for LV end-diastolic volume (4.9mL vs. 6.7), end-systolic volume (3.1mL vs. 5.2), and ejection fraction (2.6\% vs. 3.7); and much tighter limits of agreement, which were already within inter-rater variability for non-contrast echo. These results demonstrate the benefits of BO for echocardiography segmentation over a recent state-of-the-art framework, although validation using large-scale independent clinical data is required.
\end{abstract}

\comment{

}


\section{Introduction}
\label{sec:intro}

Echocardiography is the most frequently used non-invasive imaging modality for the quantification of heart structure and function~\cite{lang2015recommendations}. However, such quantification requires precise annotations of the key cardiac structures, including the left ventricular endocardium ($\text{LV}_{endo}$), epicardium ($\text{LV}_{epi}$), and left atrium ($\text{LA}$). Manual annotation is high-cost and prone to inter-rater variability, which has motivated the development of automated segmentation methods, most recently using convolutional neural networks~\cite{zhang_CIRC18_rahulStuff}.

The U-net architecture \cite{ronneberger_MICCAI15_UNET} is well-known for its capacity to learn abstract features and complete a wide range of medical image segmentation tasks. Within the domain of echocardiography segmentation, LeClerc et al.~\cite{leclerc_TMI19_CAMUS_Dataset} have demonstrated the relative efficacy of U-net models on their large annotated CAMUS dataset. 

Among others who have followed this work~\cite{wei_MICCAI20_colearningVideoSeg}, Stough et al.~\cite{stough_SPIE20_2dEchoSeg} developed a U-net variant and claimed state-of-the-art results on CAMUS. However, the hyperparameters used in this framework were found heuristically and through small-scale grid search, given the high cost of training such a deep fully convolutional neural network. In this work we leverage Bayesian Optimization (BO) and distributed computing to efficiently search the hyperparameter space, resulting in significant improvements in multi-structure echocardiography segmentation.

Bayesian optimization provides both theoretical guarantees in optimizing expensive black-box functions~\cite{bull2011convergence}, and has been proven to work well in practice~\cite{snoek2012practical}. BO is especially valuable when the observed objective is gradient-free, expensive to evaluate, and has fewer than 20 parameters to be optimized~\cite{frazier2018tutorial}. BO utilizes Gaussian Processes to gain insights into the objective function’s probability distribution, and uses an acquisition function on top of this distribution to determine the next set of hyperparameters to be tested.

\comment{

}


\section{METHODS}
\label{sec:method}  

\begin{figure*}[t]
   \begin{center}

    \includegraphics[width=0.95\textwidth]{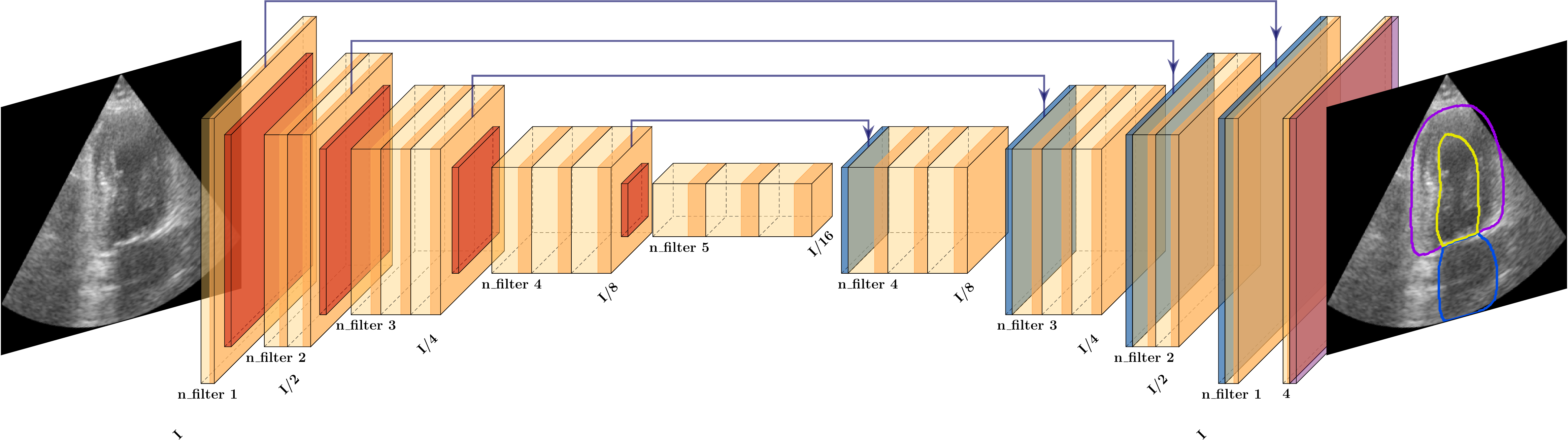}

   \end{center}
   \caption[Convolutional neural network architecture] 
   { \label{fig:CNN_arch}
CNN architecture for multi-structure segmentation of echocardiography~\cite{stough_SPIE20_2dEchoSeg}. Each block incorporates chained convolution, group normalization, and non-linear activation. During BO optimization, architectural hyperparameters include the number of output filters (n\_filter), and whether to use group or batch normalization.} 
\vspace{-8pt} 
   \end{figure*}

The model developed in~\cite{stough_SPIE20_2dEchoSeg} is an encoder-decoder style network incorporating additive skip connections and group normalization (Fig.~\ref{fig:CNN_arch}). We optimize on a variety of architectural and training-related hyperparameters, as shown in Table~\ref{fig:Hyperparameter_table}.

We then observe a noisy objective function $f(\x)$ that evaluates the performance of the model on a given candidate, or set of hyperparameters $\x \in H$: $y = f(\x) + \epsilon$, where $y$ is the observed value, and $\epsilon \sim \mathcal{N}(0, \sigma^2)$, in which $\sigma$ is given. The known performances on initial candidates will be fed into BO to find candidates $\x$ that maximize $f(\x):$ $\mathrm{argmax}_x\ f(\x)$.

\subsection{Objective Functions}
\label{ssec:objective}


\begin{table}
\centering
\begin{tabular}{|c|c|c|c|c|} 
\cline{2-5}
\multicolumn{1}{l|}{}                    & Hyperparameter                                                & Range         & \cite{stough_SPIE20_2dEchoSeg} & AP2,4 MV          \\ 
\hline
\multirow{7}{*}{\rotatebox[origin=c]{90}{Architectural}} & n\_filter \#1                                                & {[}16, 32]    & 32    & 21, 19      \\
                                         & n\_filter \#2                                                & {[}57, 128]   & 64    & 94, 77      \\
                                         & n\_filter \#2                                                & {[}153, 256]  & 128   & 225, 157    \\
                                         & n\_filter \#4                                                & {[}281, 512]  & 256   & 427, 490    \\
                                         & n\_filter \#5                                                & {[}537, 1024] & 512   & 811, 915    \\
                                         & \begin{tabular}[c]{@{}c@{}}group vs. batch\end{tabular} & {[}0, 1]      & 0     & 1, 1        \\
                                         & num. groups                                              & {[}2, 24]     & 16    &             \\ 
\hline
\multirow{3}{*}{\rotatebox[origin=c]{90}{Training}}      & lg learning rate                                             & {[}-9, 2]     & -6.22 & -8.0, -7.6  \\
                                         & lg weight decay                                              & {[}-9, -2]    & -13.8 & -7.9, -8.7  \\
                                         & batch size                                                    & {[}2, 10]     & 16    & 7, 6        \\
\hline
\end{tabular} \caption[Hyperparameter ranges] 
  { \label{fig:Hyperparameter_table}
Table of hyperparameters optimized with prior published and BO-optimal settings. AP2,4 denotes apical 2-chamber and 4-chamber view.} 
\vspace{-5pt}
\end{table}

\begin{table*}[tp]
\centering
\begin{tabular}{|l|l|l|l|l|}
\cline{1-5}
                     & Label / Score & ${\rm D}_{mean}$                              & ED ${\rm D}_{mean}$ $\rpm$ $\sigma$                                          & \multicolumn{1}{l|}{ES ${\rm D}_{mean}$ $\rpm$ $\sigma$}                         \\ 
\cline{1-5}
{AP2} & ${\rm LV}_{endo}$  & \begin{tabular}[c]{@{}l@{}}\textbf{0.950}(0.921)\\\end{tabular} & \textbf{0.960}(0.937) \rpm  \textbf{0.015}(0.034) & \textbf{0.941}(0.905) \rpm  \textbf{0.033}(0.059)    \\
                     & ${\rm LV}_{epi}$   & \textbf{0.966}(0.950)                                           & \textbf{0.968}(0.953) \rpm  \textbf{0.012}(0.024) & \textbf{0.964}(0.947) \rpm  \textbf{0.014}(0.028)    \\
                     & LA            & \textbf{0.934}(0.879)                                           & \textbf{0.929}(0.857)  \rpm~\textbf{0.038}(0.131) & \textbf{0.939}(0.901) \rpm \textbf{0.028}(0.081)    \\ 
\cline{1-5}
{AP4} & ${\rm LV}_{endo}$  & \textbf{0.954}(0.935)                                           & \textbf{0.962}(0.946) \rpm \textbf{0.016}(0.023) & \textbf{0.945}(0.924) \rpm \textbf{0.027}(0.039)    \\
                     & ${\rm LV}_{epi}$   & \textbf{0.969}(0.958)                                           & \textbf{0.971}(0.961) \rpm  \textbf{0.010}(0.015) & \textbf{0.967}(0.955) \rpm  \textbf{0.011}(0.016)    \\
                     & LA            & \textbf{0.935}(0.910)                                           & \textbf{0.924}(0.890) \rpm  \textbf{0.042}(0.071) & \textbf{0.947}(0.931) \rpm  \textbf{0.025}(0.032) \\
\cline{1-5}
\end{tabular}
\caption[Dices] 
   { \label{fig:Dices_table}
Dice overlaps for the MV optimal candidate shown against the published candidate~\cite{stough_SPIE20_2dEchoSeg} (in parentheses) on the evaluation set (136 patients).} 
\vspace{-8pt} 
\end{table*}

We score the goodness of candidate $\x$ through mean test loss in a 5-fold cross validation setting with $N$ total images. Let $E_j$ be test fold $j$, $h_\x^{j}$ be the model trained using $\x$ with fold $j$, $I$ be an image in $E_j$, and $L$ be the Cross Entropy loss from the resulting segmentation $h_\x^{j}(I)$. Since we're maximizing the objective function, we invert this test loss. We call this objective Mean Validation loss (MV), which is observed through:
\[y = o_{MV}(\x) = 1 - \frac{1}{N}\sum_{j=1}^{5} \sum_{I\in E_j} L[h_\x^{j}(I)] \qquad  \]


\subsection{Bayesian Optimization} 
\label{ssec:bayesopt}
Implemented with BoTorch \cite{balandat2019botorch}, we utilize heteroskedastic Gaussian Processes (GP), which wrap another GP to model changing objective noise. An additional GPU constraint is modelled with a fixed noise GP. All GPs use the default Matérn 5/2 kernel. Based on these GPs, the acquisition function Noisy Expected Improvement (NEI) \cite{letham2019constrained} is applied on both objective and constraint in scoring candidates.

In this setting, we assume that the observed objective value is corrupted by white noise $\epsilon \sim \mathcal{N}(0, \sigma^2)$, where $\sigma^2$ is set to be the variance of validation loss. Suppose we use Expected Improvement (EI) on the observed objective, we need to know $f^*$, or known best true objective, which is not apparent. This is due to our assumption that only the noisy observed value $y$ is known. To deal with this drawback, Letham et al.~\cite{letham2019constrained} sample multiple  random ``imaginary" instances of $[f(\x_1), \cdots, f(\x_n)]\ |\ D_f \sim \mathcal{N}(\mu_f, \Sigma_f)$, where $D_f = \{\x_i, y_i, \sigma^2_{i}\}_{i=1}^n$ are known values and uncertainty estimates of the objective. Each instance $[f(\x_1), \cdots, f(\x_n)]$ is then used to fit a predetermined $k$ number of different GP models $\mathcal{M}_1, \cdots, \mathcal{M}_k$, which offer insights into different stochastic scenarios. Additionally, a constraint is added to keep GPU usage in check, thus ensuring the worker machines are capable of handling the objective evaluation process. Similar to the objective, multiple ``imaginary" instances of constraint are sampled and fitted to multiple GPs $\mathcal{M}_1^\prime, \cdots, \mathcal{M}_k^\prime$. Each output on a potential candidate of the objective GPs  $f(\x)$ and constraint GPs $c(\x)$ is combined as follows:
\[Weighted\ Objective\ of\ \x = f(\x) \Big(1 - \frac{1}{1+e^{-c(\x)}} \Big)\]
We then calculate EI for the weighted objective of each pair $[\mathcal{M}_i, \mathcal{M}_i^\prime]$. After that, we average all EI values of a particular $\x$ to get the final NEI value.

\comment{

}


\section{Experimental Results}
\label{sec:results}  

The CAMUS dataset consists of 450 patients, with both apical two and four chamber views (AP2/AP4) per patient, and annotated end-diastolic/end-systolic (ED/ES) phases per view, totalling 1800 echocardiographic frames and corresponding label masks (LV\textsubscript{endo}, LV\textsubscript{epi}, LA, background). Additional information for each patient includes age, sex, and reported ED/ES LV\textsubscript{endo} volumes and ejection fraction (EF), along with the observed image quality for each view. Annotations for 50 additional patients are left out publicly as part of a challenge set~\cite{leclerc_TMI19_CAMUS_ChallengeSite}. 

We initially leave out $\sim$30\% (N = 136) of patients for final evaluation. The remaining 70\% (N = 314) are then partitioned for 5-fold cross validation with training, validation, and test splits. As in~\cite{leclerc_TMI19_CAMUS_Dataset}, all splits are stratified on both patient EF range ($\leq 45\%,\ \geq 55\%,$ else) and reported image quality. 

In training candidate $\x$, the model weights are saved according to performance on the validation splits, while the BO objective is computed against the associated test splits. As in~\cite{stough_SPIE20_2dEchoSeg}, on-the-fly data augmentation is used, including intensity windowing, slight rotation about the transducer point, and additive Gaussian noise. Training is continued to convergence using a standard scheduler that reduces learning rate on a plateau in validation loss.

\begin{figure}[ht]
   \begin{center}
    \includegraphics[width=0.48\textwidth]{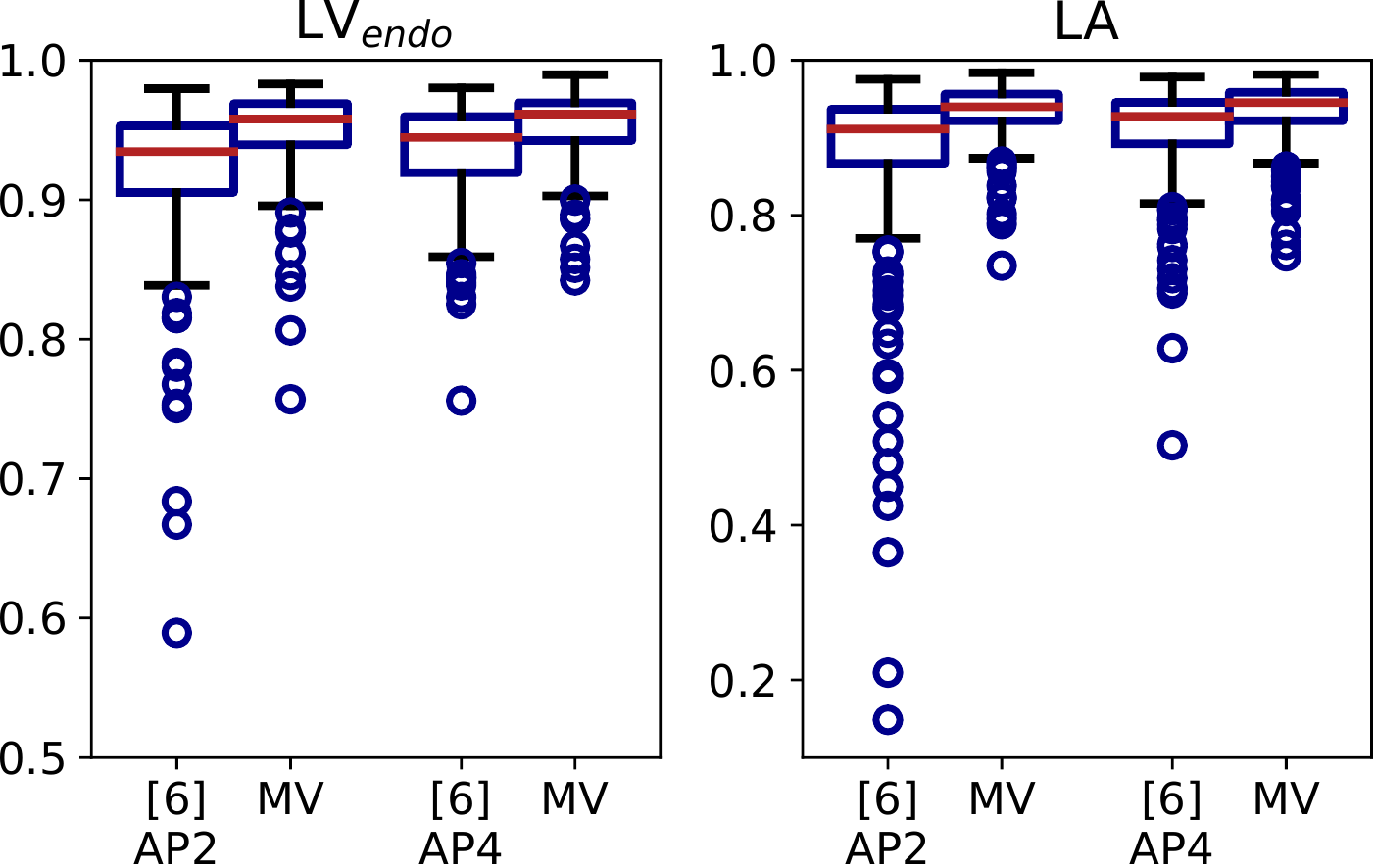}
   \end{center}
   \caption[Boxplot] 
   { \label{fig:LVendo_boxplot}
Box plots of $\text{LV}_{endo}$ and LA Dice performance. The MV optimum shows significant improvement over~\cite{stough_SPIE20_2dEchoSeg} in median and outlier performance on both structures.} 
\vspace{-5pt} 
   \end{figure}
   
\begin{figure}[h]
   \begin{center}

    \includegraphics[width=0.48\textwidth]{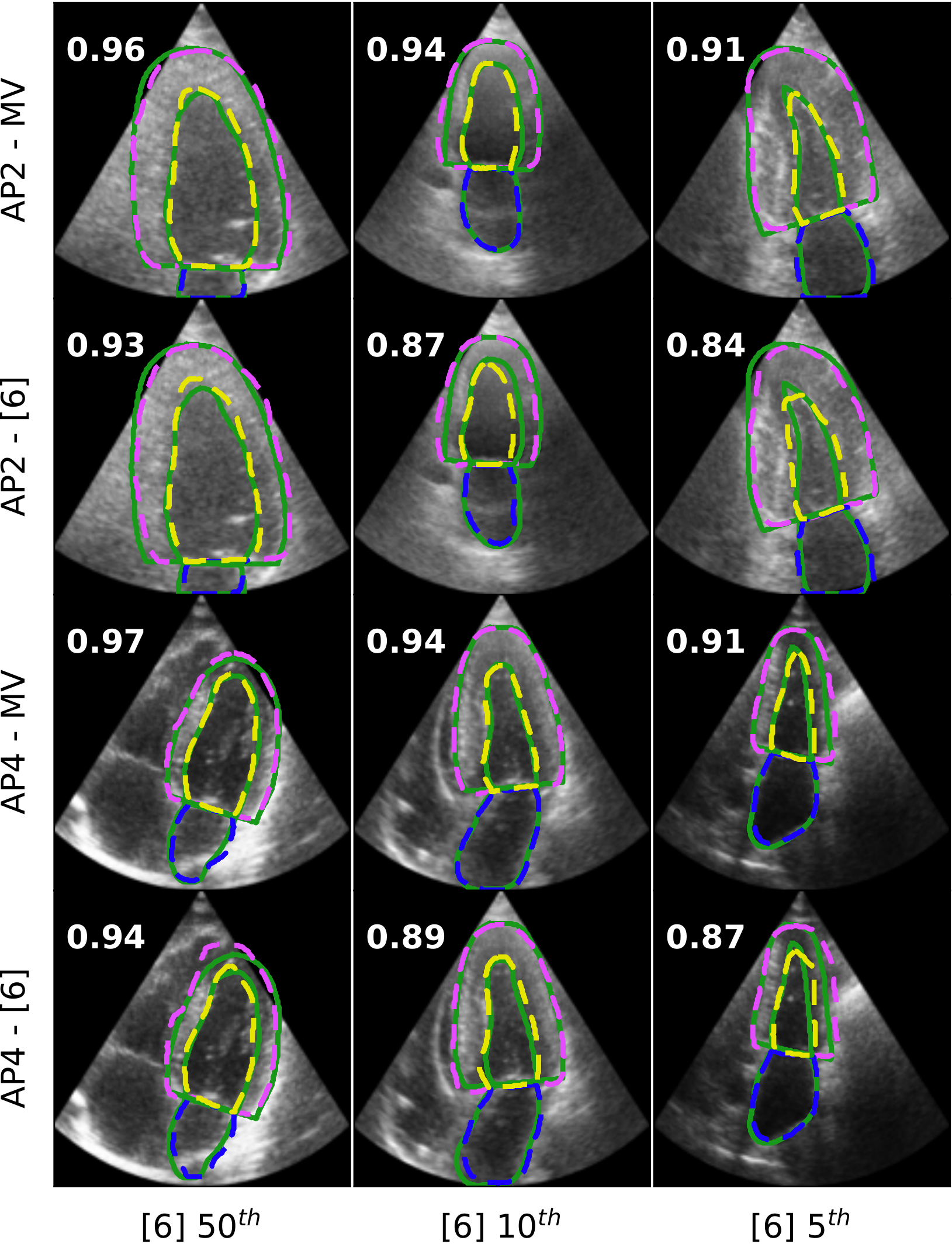}

   \end{center}
   \caption[Echosegs] 
   { \label{fig:Echo_segs} Segmentation performance of MV and published \cite{stough_SPIE20_2dEchoSeg} candidates on $50^{th},\ 10^{th}$ and $5^{th}$ $\textrm{LV}_{endo}$ Dice score percentile for AP2 and AP4 views of \cite{stough_SPIE20_2dEchoSeg}. Green contour denotes manual segmentation.} 
\vspace{-10pt} 
\end{figure}

We run BO asynchronously in a distributed environment in which each node runs a single GeForce RTX 2080 Ti. We run 100 candidates for each of AP2 and AP4 views, resulting in two best candidates. In segmenting the 30\% evaluation set with a particular optimal candidate, we accumulate the outputs from all five folds to obtain an ensemble result.

\begin{figure*}[tp]
   \begin{center}

    \includegraphics[width=\textwidth]{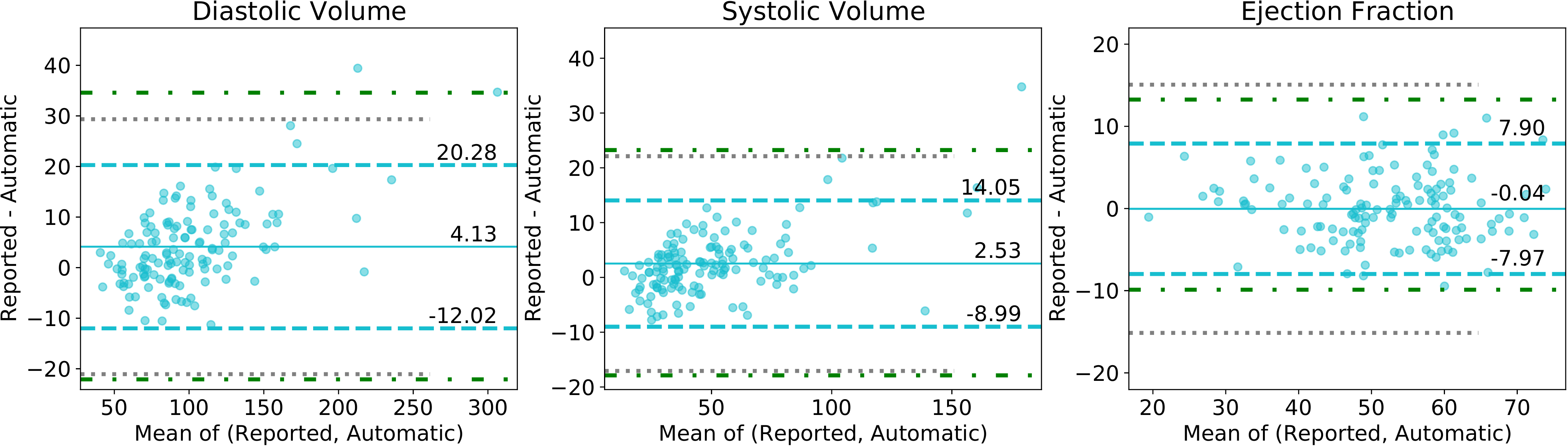}

   \end{center}
   \caption[Bland-Altman plot] 
   { \label{fig:BA_plot}
Bland-Altman plots comparing the optimal MV candidate against manual annotations for $\textrm{LV}_{endo}$ volumes and ejection fraction on the evaluation set, in light blue. Additional limits of agreement are shown for both the published candidate~\cite{stough_SPIE20_2dEchoSeg} on the same data (green, dashdotted) and previously reported inter-observer variability for 2D echocardiography~\cite{wood_Echo14_highInterRaterVariability} (gray, dotted). Compared to~\cite{stough_SPIE20_2dEchoSeg}, we report bias$ \rpm 1.96\sigma$ of $4.13\textrm{mL}\rpm16.15$ (vs $6.25\textrm{mL}\rpm 28.34$) for ED volume, $2.53\textrm{mL}\rpm11.52$ (vs $2.69\textrm{mL}\rpm20.56$) for ES volume, and $-0.04\textrm{\%}\rpm7.94$ (vs $1.69\textrm{\%}\rpm11.57$) for EF.} 
\vspace{-8pt} 
   \end{figure*} 

Table~\ref{fig:Dices_table} directly compares the MV optimal candidate to the previously published candidate~\cite{stough_SPIE20_2dEchoSeg} through Dice. The MV optimal candidate shows improved agreement with manual annotation for all views, structures, and phases. Figure~\ref{fig:LVendo_boxplot} provides additional context, showing greatly improved outlier performance of the MV optimal candidate, particularly for the relatively small and thus difficult left atrium. Figure~\ref{fig:Echo_segs} shows representative median and poor segmentation results.

We further derive the LV volume measurements and EF using the Simpson's modified biplane method and the corresponding AP2 and AP4 views for each patient. Compared to~\cite{stough_SPIE20_2dEchoSeg}, the MV optimal candidate obtains significantly smaller biases and narrower limits of agreement with reported clinical indices. We report median absolute errors of 
4.9mL (vs. 6.7mL) for ED volume, 3.1mL (vs. 5.2mL) for ES volume, and 2.6\% (vs. 3.7\%) for EF (Fig.~\ref{fig:BA_plot}).

\comment{

}


\subsection{Additional Validation}

Relative to~\cite{stough_SPIE20_2dEchoSeg}, the optimal hyperparameter sets are characterized by smaller learning rates along with deeper feature maps and thus more trainable parameters (40M vs 13M, Table~\ref{fig:Hyperparameter_table}). Along with the significant performance improvements observed comes the concern that there may be overfitting to the relatively consistent and artifact-free CAMUS images, though care was taken to separate training, BO scoring, and evaluation data. 

To assess the potential of overfitting, we additionally evaluate the generalizability of our CAMUS-trained MV optimal candidate on the large EchoNet-Dynamic clinical dataset from Stanford~\cite{ouyang_Nature2020_echoVideoSeg}. The test set contains 1276 echocardiogram videos in AP4 view only, each with single annotated ED and ES frames. $LV_{endo}$ is the only annotated structure. In segmenting such a frame, as elsewhere, we accumulate the outputs from all five folds to obtain an ensemble result. We obtain median Dice overlaps of 0.921/0.895 on ED/ES frames, only slightly reduced from models trained on the dataset itself~\cite{ouyang_Nature2020_echoVideoSeg}. These results also represent significant improvement over~\cite{stough_SPIE20_2dEchoSeg} via a paired Wilcoxon signed-rank test~\cite{pratt_WilcoxonTest}. 

Lastly, we evaluate the MV optimal candidate on the left-out 50 patients of the CAMUS challenge set~\cite{leclerc_TMI19_CAMUS_ChallengeSite}. We obtain mean ED/ES overlaps of 0.948/0.928 on $LV_{endo}$, 0.962/0.955 on $LV_{epi}$, and 0.899/0.932 on $LA$. These results are reduced versus our own left-out 30\% evaluation set (see Table~\ref{fig:Dices_table}), though consistent with prior results on the challenge set itself~\cite{leclerc_TMI19_CAMUS_Dataset, wei_MICCAI20_colearningVideoSeg}.

\comment{

}



\section{Conclusion}
\label{sec:conclusion}

In this work, we have utilized Bayesian Optimization to significantly improve upon recent state-of-the-art multi-structure segmentation in echocardiography. In a fair comparison, the optimal candidate boasts tighter limits of agreement and vastly improved outlier performance. The potential absence of catastrophic failures makes more feasible limited auditing in future large-scale historical analyses.

Model performance further generalized to a large independent clinical dataset~\cite{ouyang_Nature2020_echoVideoSeg}, providing even putative multi-structure segmentation in the absence of manual annotations. We must continue to assess generalizability to other historical clinical data, which often features larger variability in acquisition settings and image quality, and even burned-in view and patient information that is more common in the clinic.

\comment {
in the absence of manual annotations - corroboration?
}



\bibliographystyle{IEEEbib}
\bibliography{echo_research}

\begin{thebibliography}{10}

\bibitem{lang2015recommendations}
Roberto~M Lang, Luigi~P Badano, Victor Mor-Avi, Jonathan Afilalo, et~al.,
\newblock ``Recommendations for cardiac chamber quantification by
  echocardiography in adults...,''
\newblock {\em European Heart Journal-Cardiovascular Imaging}, vol. 16, no. 3,
  pp. 233--271, 2015.

\bibitem{zhang_CIRC18_rahulStuff}
Jeffrey Zhang, Sravani Gajjala, Pulkit Agrawal, et~al.,
\newblock ``Fully automated echocardiogram interpretation in clinical
  practice,''
\newblock {\em Circulation}, vol. 136, no. 16, pp. 1623--1635, 2018,
\newblock \url{https://doi.org/10.1161/CIRCULATIONAHA.118.034338}.

\bibitem{ronneberger_MICCAI15_UNET}
Olaf Ronneberger, Philipp Fischer, and Thomas Brox,
\newblock ``U-net: Convolutional networks for biomedical image segmentation,''
\newblock 2015, pp. 234--241,
\newblock \url{https://doi.org/10.1007/978-3-319-24574-4_28}.

\bibitem{leclerc_TMI19_CAMUS_Dataset}
Sarah Leclerc, Erik Smistad, Jo{\=a}o Pedrosa, et~al.,
\newblock ``Deep learning for segmentation using an open large-scale dataset in
  2d echocardiography,''
\newblock {\em IEEE Trans Med Imaging}, 2019,
\newblock \url{https://doi.org/10.1109/TMI.2019.2900516}.

\bibitem{wei_MICCAI20_colearningVideoSeg}
Hongrong Wei, Heng Cao, et~al.,
\newblock ``Temporal-consistent segmentation of echocardiography with
  co-learning from appearance and shape,''
\newblock in {\em MICCAI}, 2020,
\newblock \url{http://www.digitalimaginggroup.ca/members/Shuo/temporal.pdf}.

\bibitem{stough_SPIE20_2dEchoSeg}
Joshua~V Stough, Sushravya Raghunath, et~al.,
\newblock ``Left ventricular and atrial segmentation of 2d echocardiography
  with convolutional neural networks,''
\newblock in {\em SPIE Medical Imaging 2020: Image Processing}, 2020,
\newblock \url{https://doi.org/10.1117/12.2547375}.

\bibitem{bull2011convergence}
Adam~D Bull,
\newblock ``Convergence rates of efficient global optimization algorithms.,''
\newblock {\em Journal of Machine Learning Research}, vol. 12, no. 10, 2011.

\bibitem{snoek2012practical}
Jasper Snoek, Hugo Larochelle, and Ryan~P Adams,
\newblock ``Practical bayesian optimization of machine learning algorithms,''
\newblock in {\em Advances in neural information processing systems}, 2012, pp.
  2951--2959.

\bibitem{frazier2018tutorial}
Peter~I Frazier,
\newblock ``A tutorial on bayesian optimization,''
\newblock {\em arXiv}, 2018,
\newblock \url{https://arxiv.org/abs/1807.02811}.

\bibitem{balandat2019botorch}
Maximilian Balandat, Brian Karrer, Daniel~R Jiang, et~al.,
\newblock ``Botorch: Programmable bayesian optimization in pytorch,''
\newblock {\em arXiv preprint arXiv:1910.06403}, 2019.

\bibitem{letham2019constrained}
Benjamin Letham, Brian Karrer, et~al.,
\newblock ``Constrained bayesian optimization with noisy experiments,''
\newblock {\em Bayesian Analysis}, vol. 14, no. 2, pp. 495--519, 2019.

\bibitem{leclerc_TMI19_CAMUS_ChallengeSite}
Sarah Leclerc et~al.,
\newblock ``Camus: Cardiac acquisitions for multi-structure ultrasound
  segmentation,''
  \url{https://www.creatis.insa-lyon.fr/Challenge/camus/index.html}.

\bibitem{wood_Echo14_highInterRaterVariability}
Peter~W. Wood, Jonathan~B. Choy, Navin~C. Nanda, and Harald Becher,
\newblock ``Left ventricular ejection fraction and volumes: It depends on the
  imaging method,''
\newblock {\em Echocardiography}, vol. 31, no. 1, pp. 87--100, 2014,
\newblock \url{https://doi.org/10.1111/echo.12331}.

\bibitem{ouyang_Nature2020_echoVideoSeg}
David Ouyang, Bryan He, et~al.,
\newblock ``Video-based ai for beat-to-beat assessment of cardiac function,''
\newblock {\em Nature}, vol. 580, no. 7802, pp. 252--256, 2020,
\newblock \url{https://www.nature.com/articles/s41586-020-2145-8}.

\bibitem{pratt_WilcoxonTest}
J.W. Pratt,
\newblock ``Remarks on zeros and ties in the wilcoxon signed rank procedures,''
\newblock {\em J. American Statistical Association}, vol. 54, pp. 655--667,
  1959.

\end{thebibliography}

\section{Compliance with Ethical Standards}
This research study was conducted retrospectively using human subject data made available through registered access from the following sources: 

\begin{itemize}
    \item CAMUS: ``Cardiac Acquisitions for Multi-structure Ultrasound Segmentation,''  \\
\href{https://www.creatis.insa-lyon.fr/Challenge/camus/index.html}{https://www.creatis.insa-lyon.fr/Challenge/camus} \\
see also \\
\href{https://doi.org/10.1109/TMI.2019.2900516}{https://doi.org/10.1109/TMI.2019.2900516} \\
Ethical approval beyond citation was not required.
    
    \item EchoNet-Dynamic: ``A Large New Cardiac Motion Video Data Resource for Medical Machine Learning,'' \\
    \href{https://echonet.github.io/dynamic/}{https://echonet.github.io/dynamic/} \\
    Non-commercial research use agreement was required. Ethical approval beyond citation was not required.
\end{itemize}

\section{Acknowledgments}
This work was supported by the Ciffolillo Healthcare Technology Inventors Program through Bucknell University. The authors additionally acknowledge the Bucknell Geisinger Research Initiative.

\end{document}